%% file: main.tex
\documentclass[10pt,conference]{IEEEtran} 
\IEEEoverridecommandlockouts
\usepackage{acronym}
\usepackage{hyperref}
\usepackage{makecell}
\usepackage{pifont}
\usepackage{cite}
\usepackage{amsmath,amssymb,amsfonts}
\usepackage{algorithmic}
\usepackage{graphicx}
\usepackage{textcomp}
\usepackage{xcolor}
\usepackage{svg}
\usepackage{subfig}
\usepackage{booktabs}
\usepackage{flushend}

\DeclareRobustCommand*{\IEEEauthorrefmark}[1]{%
  \raisebox{0pt}[0pt][0pt]{\textsuperscript{\footnotesize\ensuremath{#1}}}}
  
\def\BibTeX{{\rm B\kern-.05em{\sc i\kern-.025em b}\kern-.08em
    T\kern-.1667em\lower.7ex\hbox{E}\kern-.125emX}}

\begin{document}

\input{acronyms.tex}

\title{AtomFlow: An End-to-End FPGA-Based Control Architecture for Neutral Atom Quantum Computers\thanks{\IEEEauthorrefmark{*} Equal Contribution.\newline This work was funded by the German Federal Ministry of Research, Technology and Space (BMFTR) under the funding program Quantum Technologies - From Basic Research to Market under contract number 13N16087, as well as from the Munich Quantum Valley (MQV), which is supported by the Bavarian State Government with funds from the Hightech Agenda Bayern.}}

\author{\IEEEauthorblockN{Xiaorang Guo\IEEEauthorrefmark{*}, Jonas Winklmann\IEEEauthorrefmark{*}, Vengkeat Chea and Martin Schulz}

\IEEEauthorblockA{Chair of Computer Architecture and Parallel Systems\\TUM School of Computation, Information and Technology (CIT)\\Technical University of Munich, Garching, Germany\\
Email: \{xiaorang.guo, jonas.winklmann, vengkeat.chea, martin.w.j.schulz\}@tum.de}}

\IEEEpubid{\begin{minipage}[t]{\textwidth}\ \\[10pt]
\centering
\copyright 2026 IEEE. Personal use of this material is permitted. Permission from IEEE must be obtained for all other uses, in any current or future media, including reprinting/republishing this material for advertising or promotional purposes, creating new collective works, for resale or redistribution to servers or lists, or reuse of any copyrighted component of this work in other works.
\end{minipage}}

\maketitle

\begin{abstract}
Neutral Atom Quantum Computing (NAQC) is an emerging modality for scalable quantum computation, valued for its long coherence times and the naturally identical atomic qubits. However, one of the main drawbacks is its slow execution rate, dominated by lengthy classical processing tasks, such as fluorescence imaging, cooling, and atom rearrangement. We address this bottleneck with \textit{AtomFlow}, a field-programmable gate array (FPGA)-based control architecture that consolidates fluorescence-image analysis and a newly developed atom-rearrangement algorithm onto a single Zynq UltraScale+ device. By co-locating the two stages on the same board and emitting rearrangement moves in a streaming fashion as soon as they are computed, AtomFlow eliminates the round-trip latency of conventional host-mediated pipelines. Evaluated on a 16$\times$16 atom array, AtomFlow achieves an end-to-end latency of 25.3 ms with a first-move latency of 4 ms and an average move generation of 1 ms. Furthermore, our scalability analysis demonstrates that the architecture can readily support larger atom arrays within a single-board resource budget.

\end{abstract}


\input{chapters/introduction}
\input{chapters/backgrounds}
\input{chapters/relatedwork}
\input{chapters/architecture}
\input{chapters/evaluation}
\input{chapters/conclusion}

\bibliographystyle{IEEEtran}
\bibliography{ref.bib}

\end{document}

%% file: acronyms.tex
\acrodef{CPU}[CPU]{central processing unit}
\acrodef{FPGA}[FPGA]{field-programmable gate array}
\acrodef{ASIC}[ASIC]{application-specific integrated circuit}
\acrodef{GPU}[GPU]{graphics processing unit}
\acrodef{QPU}[QPU]{quantum processing unit}
\acrodef{PSF}[PSF]{point-spread function}
\acrodef{NISQ}[NISQ]{noisy intermediate-scale quantum}
\acrodef{PL}[PL]{programmable logic}
\acrodef{PS}[PS]{processing system}
\acrodef{DRAM}[DRAM]{Dynamic Random Access Memory}
\acrodef{DDR}[DDR]{Double Data Rate}
\acrodef{IP}[IP]{Intellectual Property}
\acrodef{MMIO}[MMIO]{Memory-Mapped I/O}
\acrodef{HLS}[HLS]{High-level Synthesis}
\acrodef{std}[std]{standard deviation}
\acrodef{HPC}[HPC]{high performance computing}
\acrodef{API}[API]{application programming interface }
\acrodef{NAQC}[NAQC]{neutral atom quantum computer}
\acrodef{NA}[NA]{neutral atom}
\acrodef{QEC}[QEC]{quantum error correction}
\acrodef{QCP}[QCP]{quantum control processor}
\acrodef{AWG}[AWG]{arbitrary waveform generator}
\acrodef{AOD}[AOD]{acousto-optic deflector}
\acrodef{RF}[RF]{radio frequency}
\acrodef{SLM}[SLM]{spatial light modulator}
\acrodef{FSM}[FSM]{Finite State Machine}

%% file: chapters/introduction.tex
\section{Introduction}
Achieving practical quantum advantage has become a central task in quantum computing~\cite{piccinelli2025quantum}, driving rapid progress across physical qubit modalities such as superconducting qubits~\cite{siddiqi2021engineering}, trapped ions~\cite{bernardini2024quantum}, photonic systems~\cite{psiquantum2025manufacturable}, and \acp{NA}~\cite{bluvstein2024logical,veroni2024optimized}. Among these, \acp{NA} are increasingly recognized as a leading candidate for large-scale quantum processors, mainly due to their long coherence times~\cite{evered2023high}, good scalability~\cite{manetsch2025tweezer}, and flexible connectivity between atoms~\cite{romao2026multiq}. However, within the control pipeline of \acp{NAQC}, the initialization and readout stages remain dominant latency bottlenecks. These stages are particularly critical when executing complex quantum programs under limited coherence-time budgets, and become even more demanding in the presence of \ac{QEC}, which requires repeated rounds of high-fidelity readout.

In \acp{NAQC}, qubit states are encoded in quantum states of individual atoms and manipulated via laser-driven operations~\cite{stade2025routing}. Prior to gate execution, atoms are loaded probabilistically into a two-dimensional array of optical traps, resulting in random occupancy states that complicate circuit mapping. Consequently, an initialization procedure is required, consisting of atom detection through fluorescence imaging, followed by rearrangement of present atoms to construct a smaller defect-free array. During program execution, measurements are also performed, both to produce final outputs and to enable conditional operations through mid-circuit feedback. In both cases, a reliable image processing mechanism is indispensable. Moreover, image processing and rearrangement are highly algorithmically intensive stages in the control pipeline, and their efficiency directly determines system-level execution latency.

\begin{figure}[tb]
   \centering
\includegraphics[page=1,width=.48\textwidth]{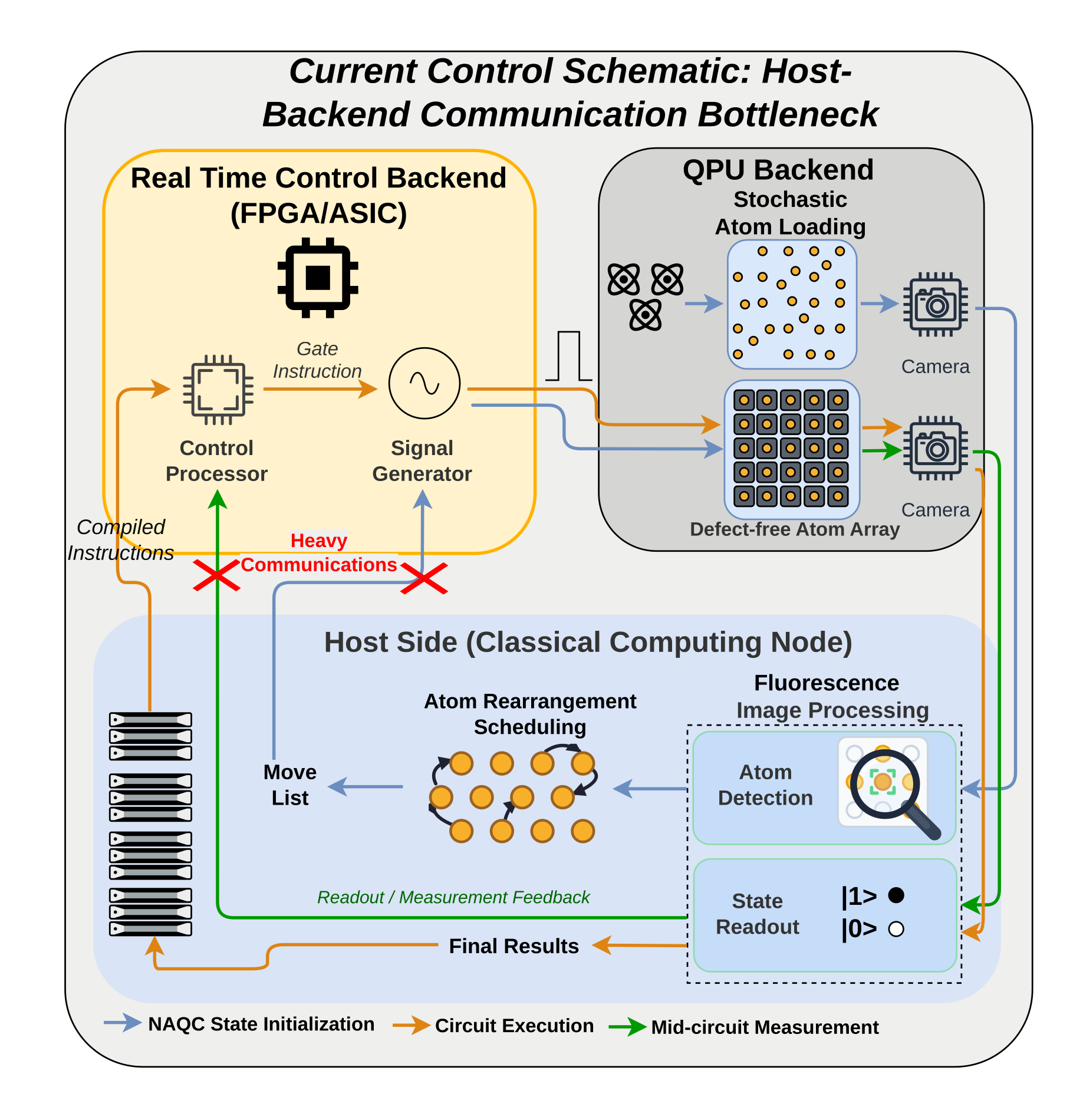}
 \caption{Abstract Representation of the current control schematic of \acp{NAQC}.}
 \label{fig:control_schematic}
\end{figure}

To date, several works have focused on software-level optimizations for these processes~\cite{Winklmann:1,winklmann2025hipars,wang2023accelerating}. Meanwhile, there is a clear trend toward implementing \acp{QCP} and \acp{AWG} on customized hardware platforms such as \acp{FPGA} and \acp{ASIC}, in order to reduce control latency and achieve tighter integration with \ac{QPU} backends~\cite{GuoHiSEP,stefanazzi2022qick,Qtenon,xu2023qubic,liu2025risc,HISQ}. Fig.~\ref{fig:control_schematic} illustrates the resulting control schematic of \acp{NAQC}, where three colored paths represent the major control flows. (1) \textit{Atom array initialization (blue path)}.
Before circuit execution, a defect-free atom array must be prepared. If atom detection and rearrangement scheduling are performed on the host side, the generated move list must be transferred to the \ac{FPGA}-based \acp{AWG} to drive \acp{AOD} for physical atom relocation. This introduces host–backend communication during every initialization round. (2) \textit{Circuit execution (orange path)}. During execution, the host compiles quantum circuits into executable instructions and transmits them to the \ac{QCP}. The \ac{QCP} handles applying qubit gates on target atoms and controls the time of \acp{AWG} generating the microwave pulses. (3) \textit{Mid-circuit measurement and feedback (green path)}.
When mid-circuit measurements occur, measurement results must be analyzed on the host side and then transmitted back to the \ac{QCP} for conditional operations. Under workloads such as \ac{QEC}, where measurement–feedback loops are frequent and latency-critical, this round-trip communication becomes particularly costly. 

In both steps of (2) and (3), the control loop inevitably crosses the boundary between the software host and the real-time control backend. These cross-boundary communications are unnecessary from a functional perspective but unavoidable under a host-centric implementation. Consequently, software-only optimizations are insufficient to satisfy stringent latency constraints, especially for \ac{QEC} workloads that demand rapid and repeated feedback. Therefore, recent works have proposed customized \ac{FPGA} implementations to accelerate image processing~\cite{jonas_date} and atom rearrangement~\cite{Guo:1}, significantly reducing processing latency. However, these efforts typically optimize the two procedures in isolation, without exploring an integrated control framework that enables an \ac{FPGA}-based solution for the end-to-end control pipeline. 

To address this limitation, we propose \textit{AtomFlow: An End-to-End FPGA-Based Control Architecture for Neutral Atom Quantum Computers} in this work. This unified framework integrates fluorescence image processing and atom rearrangement within a single \ac{FPGA}-centric pipeline, eliminating the round-trip communication overhead between off-chip host and \acp{FPGA} as shown in Fig.~\ref{fig:control_schematic}. By tuning the mode configuration, we also easily switch between initialization and readout tasks with no extra cost.

In particular, our contributions are as follows:
\begin{itemize}
    \item We adapt a parallel by-row rearrangement algorithm to \ac{FPGA} hardware via \ac{HLS}, with parameterized buffer sizing for atom array scalability.
    \item We integrate this \textit{rearrangement engine} with a previously published \ac{FPGA}-based atom detection module, exposing a unified low-latency control interface to the host.
    \item We design a streaming communication interface that delivers move data to the host as it is generated, enabling \textit{interleaved move generation and execution}.
\end{itemize}

We implement AtomFlow in \ac{HLS} and deploy it on a Xilinx Zynq UltraScale+ RFSoC~\cite{AMDrfsoc} clocked at 100 MHz. We evaluate the design across multiple atom-array configurations to verify the end-to-end workflow and characterize its scalability. On a $16\times16$ baseline, AtomFlow generates rearrangement moves at an average rate of \textbf{1.01 ms per move}, which fits within the millisecond-scale budget of many backend \ac{AWG} pulse durations.

Following this introduction, we briefly outline the background knowledge required for fluorescence imaging and atom rearrangement, followed by existing approaches and relevant adjacent works. Having done so, we introduce our system architecture and subsequently evaluate it. Finally, we summarize our findings and provide an outlook on future developments.

%% file: chapters/backgrounds.tex
\section{Background}
In order to properly grasp the requirements and constraints of the physical processes that we aim at controlling in this project, we briefly introduce the processes of fluorescence imaging and atom rearrangement, both of which are vital steps in the initialization routine of \acp{NAQC}.
\subsection{Fluorescence Image Processing}
Fluorescence image processing is a fundamental step in the operation of \acp{NAQC}, serving two critical functions: initial atom detection and final quantum-state readout. Since atoms are stochastically loaded into a two-dimensional grid of optical tweezers with a roughly 50\% probability, an initial image is typically captured and processed to identify the precise locations of defects for the preparation of the following rearrangement steps~\cite{wang2023accelerating,Guo:1}. Furthermore, at the end of the quantum circuit or during the mid-circuit measurement, qubit states can be measured through state-dependent fluorescence: all atoms in the array are illuminated with a laser; a qubit in state $|1\rangle$ will scatter photons, while a qubit in state $|0\rangle$ remains dark~\cite{mude2025enabling}. In both cases, the scattered photons are collected by a highly sensitive camera to obtain the fluorescence image. In order to minimize further latencies and heating of the atoms, exposure time should be kept to a minimum. Therefore, each atom emits only a small number of photons and the optics can only collect a fraction of them, leading to noisy detection results. Accurately processing the images requires robust processing methods, and a trade-off has to be made between the exposure time and the readout fidelity.
\subsection{Atom Rearrangement}\label{chapter:background_rearrangement}
After determining the subset of optical traps that are initially occupied, we can rearrange the present atoms using movable tweezers. In our case, we can generate a grid of such movable traps using a pair of \acp{AOD}, oriented perpendicular to each other, each being fed a number of \ac{RF} tones. In supplying several such frequencies, we can generate a 2-D grid of movable traps, which we can utilize to rearrange atom in parallel.

To execute any given move, we choose the starting rows and columns and ramp up the waveform's amplitude to increase the depth of the movable trap so it exceeds the depth of the stationary one, thereby transferring the corresponding atoms to the rearrangement grid. By altering the supplied frequencies, we can shift the rows and columns of our grid, along with any contained atoms. After arrival at their target locations, we can transfer the transported atoms back to the stationary grid by ramping down the amplitude again. The time it takes to physically execute such a move can be broken down into a constant part for transferring from and to the stationary traps, taking around 120~$\mu s$ to 800~$\mu s$ in total, and a component scaling with the move distance, where typical speeds range from 54~$\frac{\mu m}{ms}$ to 550~$\frac{\mu m}{ms}$ \cite{Gyger:1, Tian:1, bluvstein2024logical, bluvstein2022}, resulting in total execution times often exceeding 1~$ms$ per move. In light of efficient scaling requirements towards setups exceeding a thousand qubits, several libraries aiming at efficient rearrangement have emerged \cite{winklmann2025hipars, atommovr}.

\begin{figure}
    \centering
    \includegraphics[width=0.9\linewidth]{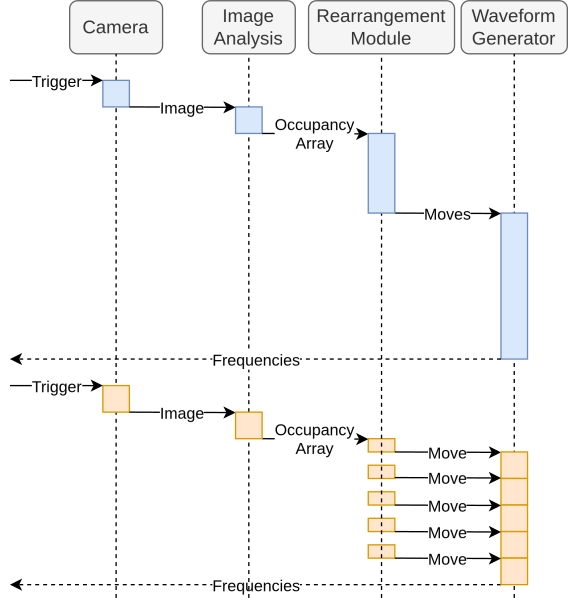}
    \caption{Sequential (blue) sequence diagram of rearrangement procedure compared to interleaved approach (yellow)}
    \label{fig:interleavingScheme}
\end{figure}

However, any algorithm that generates a complex set of moves is subject to a tradeoff between the sophistication of the generated move set and the time it takes to generate it. Usually, the run time of such an algorithm is directly added to the move execution time. To alleviate these constraints, an optimized atom rearrangement module is required, ideally tightly integrated with the waveform generation to allow for an interleaved execution, where the next move is generated while the previous one is being executed. As long as the physical execution time of each move exceeds the time required to generate the next one, this scheme, as shown in Fig.~\ref{fig:interleavingScheme}, can completely eradicate the overhead created by generating any moves other than the first.

%% file: chapters/relatedwork.tex
\section{Related Work}\label{chapter:relatedWork}
Having introduced the underlying procedures that are to be conducted, we can now investigate which solutions have already been developed and what remains to be done.

To begin with, let us discuss the general state of integrated control units for quantum computing. In recent years, it has become increasingly obvious that such control hardware is vital to the continued growth of quantum computing systems, as there is demand for specialized accelerators in various areas of application, such as readout \cite{Guo:Klinq, QubiCML,mude2025efficient}, decoding of error correction codes \cite{Maurer:1,LILLIPUT,Astreas}, and efficient quantum control processors~\cite{Guo:2023,liu2025risc,Qtenon}. While there exist approaches that can deal with several modalities already \cite{UQP}, modules specific to neutral atoms, especially for initialization and maintenance, are still lacking scalability and optimization.

Regarding the analysis of fluorescence images, there exists a multitude of approaches and custom algorithms that aim at maximizing the information that can be extracted from such images, as well as some endeavors to benchmark and compare them \cite{Tian:1, Winklmann:1}. Based on the state-reconstruction algorithm \cite{Wei:1}, which emerged as a valid contender from one of these comparative works, we ourselves have developed an \ac{FPGA}-based detection unit that, given the fluorescence image, a deconvolution kernel, and the trap locations, returns the boolean matrix describing where atoms were detected \cite{jonas_date, jonas_dac}. This will serve as the rearrangement module for this work's integration solution.

As with the rearrangement, so we have also previously investigated \ac{FPGA}-based sorting solutions \cite{Guo:1}. However, these currently lack the parallelism in the physical execution that is required to efficiently scale into the thousands of qubits. As such, we have to find an algorithm that is well-suited but yet to be adapted for the usage of \acp{FPGA}.

\begin{figure}
    \centering
    \includegraphics[width=0.95\linewidth]{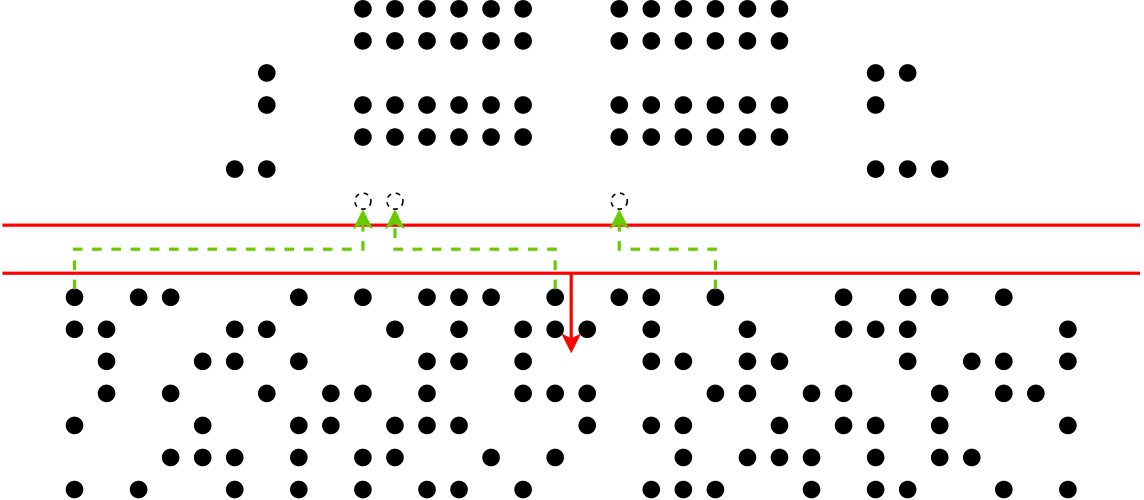}
    \caption{By-row lattice sorting scheme. The sorting channel (red) traverses the array (along the red arrow), and atoms (black circles) can move (green dashed arrows) freely within it towards their target positions (empty dashed circles).}
    \label{fig:latticeSorting}
\end{figure}

As this baseline algorithm for the rearrangement module, we use our \textit{HiPARS} sorting library \cite{winklmann2025hipars}, specifically a version of the row-by-row lattice sorting as, while not optimal for smaller tweezer setups, is robust enough to be able to deal with any considered constraints. Fig.~\ref{fig:latticeSorting} shows the idea behind this approach. In general, we employ a sorting channel that is wide enough to freely move atoms within it. Following an initial analysis step, we start by clearing the channel of its initially contained atoms, followed by a row-by-row sweep across the array. For each traversed index, we remove unusable atoms towards the outside and shift useful ones towards their target position in the desired geometry. This target geometry may be specified by the user and is usually a smaller area within the total array, in which the desired occupancy can be configured. Outside of this target region, we do not care whether a site contains an atom or not. As such, we use this unimportant area for stashing superfluous atoms that may prove useful later. If, for example, there are not enough atoms in the last rows, these parked atoms will be used to fill any remaining vacancies. This ensures that, within the target geometry, each site matches our desired target occupancy after rearrangement. In Fig.~\ref{fig:latticeSorting}, the 2$\times$6 blocks of atoms above the red sorting channel lie within the target region, while the atoms to the left and right are parked outside for later use.

\begin{figure}[tb]
  \centering
  \subfloat[2-step move]{\includegraphics[scale=0.55]{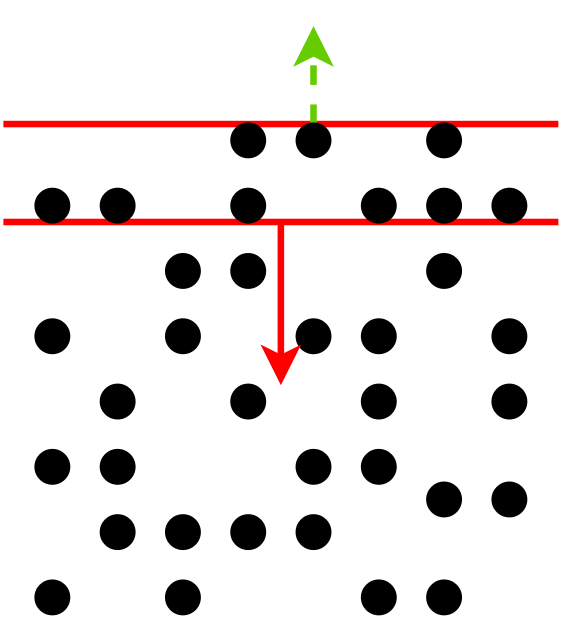}
  \label{fig:2-step}}
  \hfill
  \subfloat[3-step move]{\includegraphics[scale=0.55]{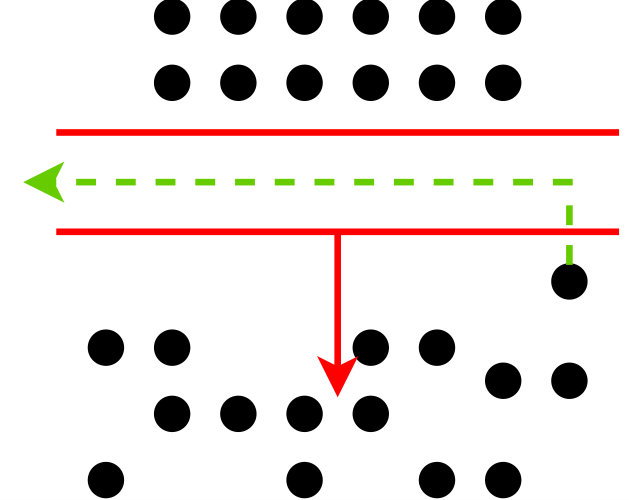}
  \label{fig:3-step}}
  \hfill
  \subfloat[4-step move]{\includegraphics[scale=0.55]{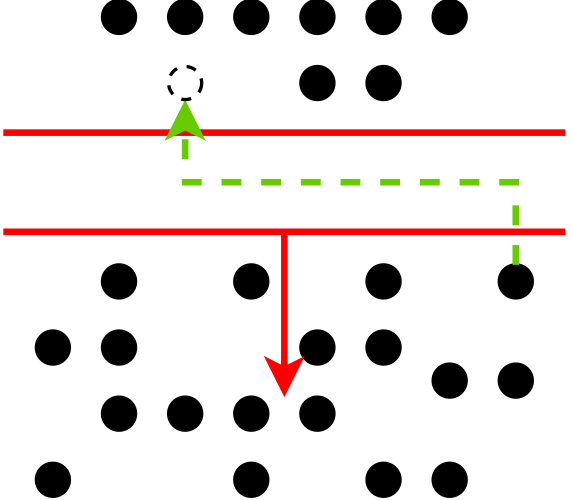}
  \label{fig:4-step}}
  \caption{Different types of moves. (a) 2-step move. Used for directly removing atoms from the sorting channel at the beginning. (b) 3-step move. Used for removing atoms through the sorting channel. (c) 4-step move. Used for moving atoms through the sorting channel towards their target positions.}
  \label{fig:move-steps}
\end{figure}

Throughout the following sections, we will occasionally mention the number of steps that a move possesses. Fig.~\ref{fig:move-steps} shows these different types of moves. Typically, each type seen in the three sub-figures is only used for one kind of action: initial clearing of the sorting channel, removal of unusable or superfluous atoms, and transportation of usable atoms towards their target locations.

%% file: chapters/architecture.tex
\section{Architecture}
In this section, we introduce the system architecture of AtomFlow and present the details of our newly proposed rearrangement accelerator. Specifically, to construct this end-to-end framework, we integrate our previously developed atom detection accelerator~\cite{jonas_date} as the \textit{Image Processing Module}, with modified interfaces. Concurrently, we introduce a newly designed \textit{Atom Rearrangement Engine} based on HiPARS~\cite{winklmann2025hipars}, coupled with a dedicated \textit{Control Interface Module} to orchestrate module interconnection. AtomFlow is open-source and available via GitHub\footnote{https://github.com/caps-tum/ATOMFLOW}.

\subsection{System Overview}
\begin{figure}[tb]
   \centering
\includegraphics[page=1,width=\linewidth]{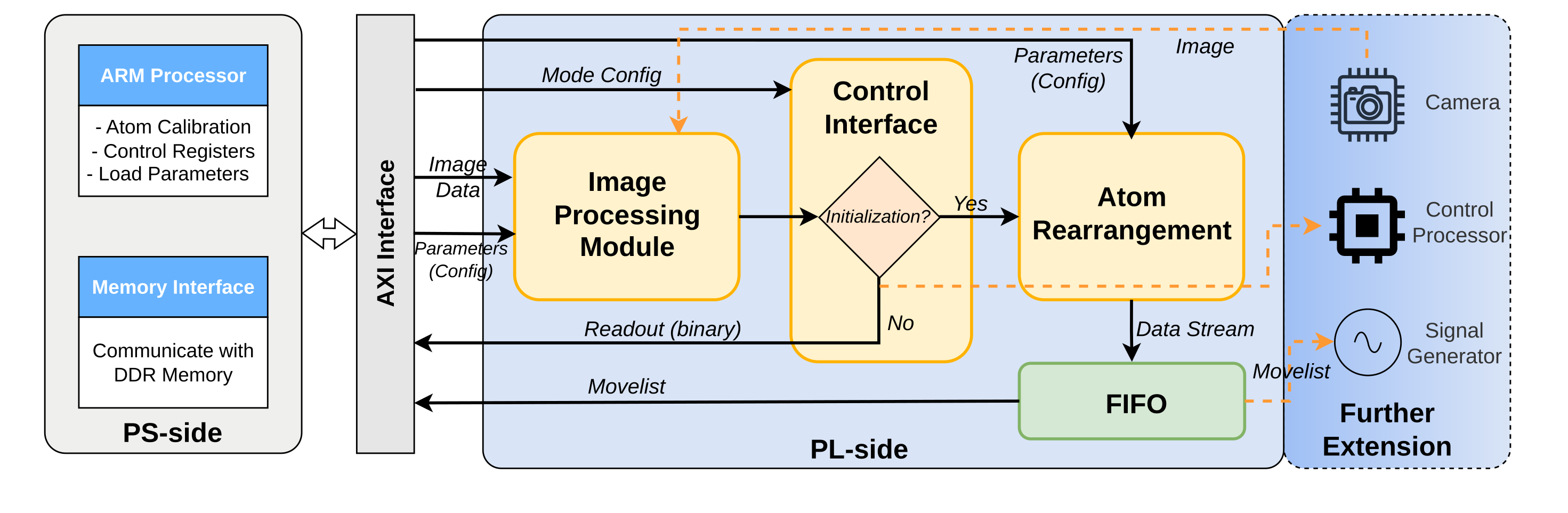}
 \caption{System architecture of AtomFlow. Solid black lines show the current software-in-the-loop implementation with the help of the \ac{PS}. Dashed orange lines indicate the planned hardware extensions, which incorporate a Camera, Signal Generator, and Control Processor, for real-time execution and mid-circuit feedback.}
 \label{fig:block_diagram}
\end{figure}

The proposed AtomFlow architecture follows a feed-forward dataflow structure as shown in Fig.~\ref{fig:block_diagram}. AtomFlow operates in two distinct modes: initialization and readout, both orchestrated by the \ac{PS} via the AXI4-Lite protocol~\cite{AXI4Lite}. At each experimental cycle, the raw fluorescence images captured by the camera are first processed by the \textit{Image Processing Module}. This module performs per-site atom occupancy reconstruction using \ac{PSF} calibration data provided by the ARM processor on the \ac{PS} side.

If the mode is configured as initialization, the \textit{Control Interface Module} binarizes the reconstructed emission values against a pre-set threshold to produce an exact atom occupancy map. This resulting binary map is then passed to the \textit{Atom Rearrangement Engine}, which computes parallel atom moves and streams them move-by-move through an AXI4-Stream FIFO to the \ac{PS} for output verification. Conversely, in the readout mode, the reconstructed atom emissions are returned directly to the \ac{PS} for qubit state evaluation, enabling essential mid-circuit measurements during quantum circuit execution.

It should be noted that, in the current implementation (indicated by solid lines in Fig.~\ref{fig:block_diagram}), the \ac{PS} is utilized to orchestrate data movement and perform result analysis. However, in the ultimate target lab setup (indicated by orange dashed lines), the pipeline will be mainly hardware-focused, with parameters and calibrations remaining on the \ac{PS} side: fluorescence images will be directly streamed from the camera into the \textit{Image Processing Module}, the output stream from the FIFO will directly feed the \ac{FPGA}-based signal generators to drive the \acp{AOD}, and during readout, the reconstructed emissions will be analyzed directly on the \ac{FPGA}-based control processor for fast feedback purpose.

Furthermore, the \ac{PS} is responsible for three primary tasks. First, it performs initial system calibration, which includes detecting atom positions and grid orientation, calculating the inverse kernel, and determining the binarization threshold. Because this calibration is performed offline and remains valid across multiple experimental runs, it only requires re-execution when the physical optical setup changes. Second, the \ac{PS} manages parameter and register configuration. To support the dynamic reconfigurability of both the \textit{Image Processing Module} and the \textit{Atom Rearrangement Engine}, the \ac{PS} handles the transmission of run-time variables, such as calibrated atom locations, binarization thresholds, the \ac{PSF} kernel, and the target arrangement zone, while also configuring their respective control registers. Last, it needs to fetch the atom-array data from memory and execute the final result analysis.

\subsection{FPGA-based Rearrangement Engine}

\begin{figure}[tb]
   \centering
\includegraphics[page=1,width=\linewidth]{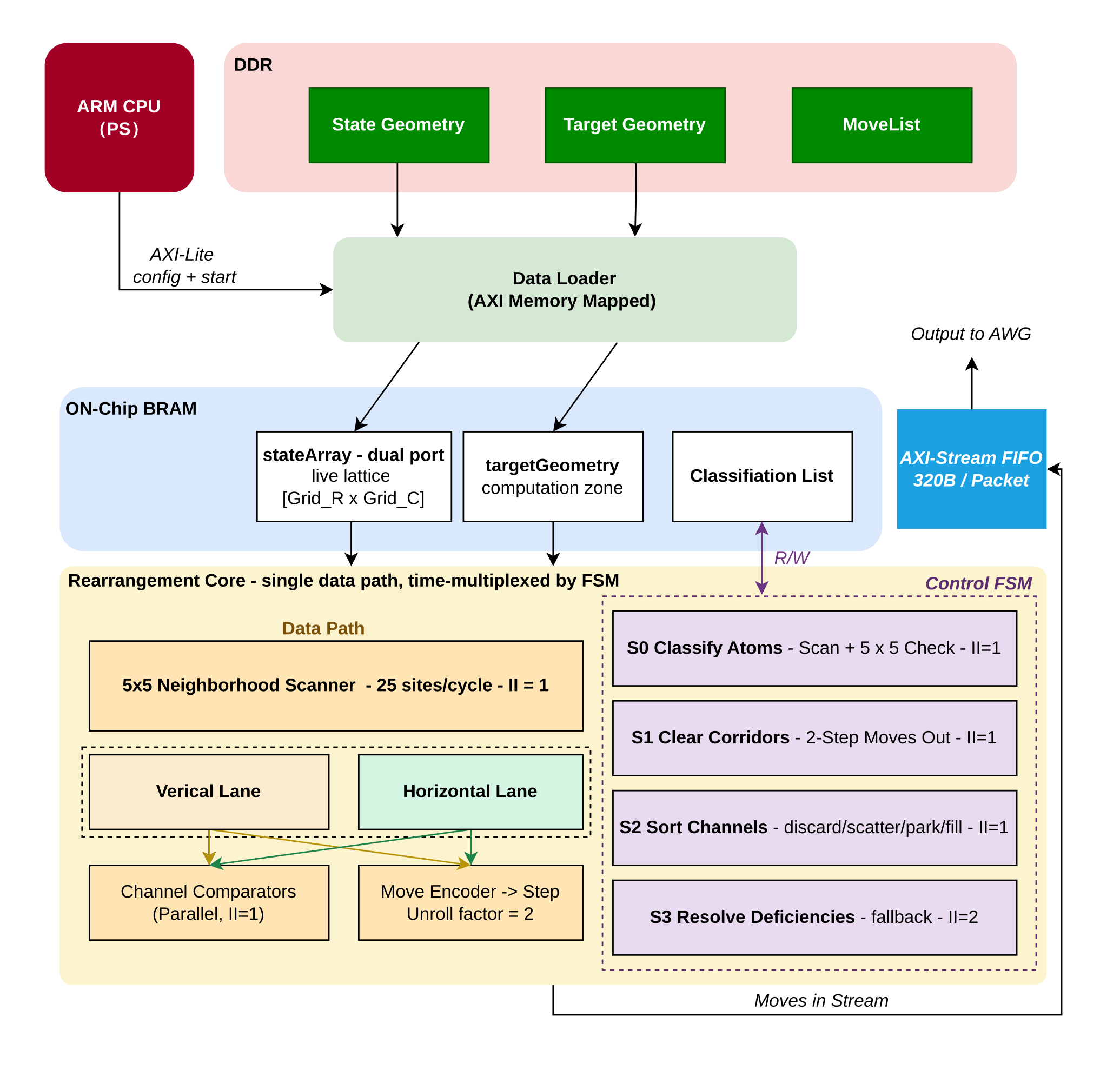}
 \caption{High-level architecture of the FPGA-based Atom Rearrangement Engine and its associated data flow.}
 \label{fig:atom-rearrangement-kernel}
\end{figure}

Fig.~\ref{fig:atom-rearrangement-kernel} shows the detailed architecture of the FPGA-based rearrangement kernel and its data flow. The design is structured as a hardware-accelerated load--compute--stream pipeline that fuses memory transfer, on-chip processing, and result delivery into a single accelerator. As illustrated at the top of the figure, the \ac{PS} (ARM CPU) uses the AXI-Lite control interface to supply the kernel with computation-zone boundaries, array dimensions, and base addresses, subsequently triggering execution via the standard \texttt{ap\_start}/\texttt{ap\_done} handshake. The initial site-occupation array and the target geometry are loaded from external DDR memory through the AXI Memory-Mapped Data Loader, while the generated moves are emitted on the fly as an AXI-Stream that drives the downstream \ac{AWG}, as indicated by the ``Output to AWG'' arrow on the right of the figure. However, in this work, we stream the output back to the \ac{PS} for analysis.

The kernel operates on a two-dimensional binary array of up to $256\times256$ sites, where each cell encodes the presence or absence of an atom. The \emph{state geometry} captures the current occupancy pattern, while the \emph{target geometry} encodes the desired final arrangement; the \emph{computation zone} is the rectangular sub-region of the lattice delimited by row and column boundary indices, within which all rearrangement decisions are made, as we described in Sec.~\ref{chapter:background_rearrangement}.

\paragraph{Input validation and data loading}
Execution begins with a validation stage that checks (i)~the lattice dimensions are within hardware limits and (ii)~the computation zone lies strictly inside the lattice. The state and target geometries are then loaded into separate on-chip BRAM buffers (\texttt{stateArray} and \texttt{targetGeometry} in the BRAM panel of Fig.~\ref{fig:atom-rearrangement-kernel}); although stored as flat byte streams in DDR, both are reshaped on the fly into 2-D grids during the burst transfer. The \texttt{stateArray} buffer is bound to a dual-port BRAM so that the live occupancy data can be both read by the classifier and written by the move executor in the same cycle, which is what makes the downstream pipelines achievable at initiation interval $\mathrm{II}=1$.

\paragraph{Rearrangement core overview}
The rearrangement core is driven by a four-state \ac{FSM} clocked at 100~MHz. As annotated in the right column of Fig.~\ref{fig:atom-rearrangement-kernel}, the four \ac{FSM} states (S0--S3) execute sequentially and share a single physical datapath that features time-multiplexing. Two HLS-level optimizations are central to the core's throughput. First, vertical and horizontal traversals are emitted as separate hardware instances, producing the two specialized lanes shown in the datapath panel of Fig.~\ref{fig:atom-rearrangement-kernel} (yellow ``Vertical Lane'' and green ``Horizontal Lane''). Both lanes feed the shared Channel Comparators and Move Encoder downstream, eliminating axis-selection multiplexers from the steady-state critical path. Second, every inner scan, scatter, and compaction loop is annotated with \texttt{\#pragma HLS PIPELINE II=1}, sustaining one element per cycle through the comparator and move-encoder stages, as shown by the \texttt{II=1} tags on states S0--S2 and the \texttt{II=2} tag on S3 in the figure. The bidirectional purple ``R/W'' arrow connecting to the rearrangement core indicates simultaneous read and write operations every clock cycle. Supporting one read and one write per cycle is critical, as it enables in-place compaction at every stage.

\subsubsection*{Atom Classification (S0)}
The classifier sweeps the lattice site by site and builds three per-index lists in dual-port BRAM. A site that lies within the target region and that should be occupied after rearrangement is appended to \texttt{targetSites}. An occupied site is tested against a $5\times5$ neighborhood mask: if spacing conditions are met, i.e., if no other atoms are so close as to prohibit reliable separation, it is appended to \texttt{usableAtoms}, otherwise to \texttt{unusableAtoms}. The mask evaluates 25 sites per cycle, and the lookup array backing the mask is partitioned with \texttt{\#pragma HLS ARRAY\_PARTITION cyclic factor=2} so that the mask reads do not serialize on a single BRAM port. The output of S0 is reused by every subsequent stage, which is the main reason classification is hoisted out as the first FSM state.

\subsubsection*{Corridor Clearing (S1)}
Before atoms can be relocated between indices, a transit corridor must be opened near the lattice boundary, which we call the sorting channel. Indices are processed sequentially; within each one, occupied corridor sites are gathered into a fixed-size selection buffer to match the parallelism of the move encoder, and are then emitted as a single two-step \emph{ParallelMove} that ejects all selected atoms to an out-of-lattice destination. Each move is applied immediately to the live state buffer so that subsequent classifications see the up-to-date occupancy. Note that corridor clearing does not consume the S0 classification lists; it operates directly on the state array and prepares the substrate that S2 and S3 depend on. The emitted moves are forwarded straight onto the output stream (the ``Moves in Stream'' path at the bottom of Fig.~\ref{fig:atom-rearrangement-kernel}).

\subsubsection*{Channel Sorting (S2)}
This is the main rearrangement phase. For each index, we consume atoms from its \texttt{unusableAtoms} and \texttt{usableAtoms} list and fill the elements of the relevant \texttt{targetSites}. Notable, we increment the target index only once all current target sites are filled in order to avoid gaps in central indices. Per index, intermediate \emph{discard}, \emph{scatter}, \emph{park}, and \emph{fill} operations may take place that are resolved against the same lists in place:

\begin{itemize}
  \item \textbf{Discard}: Surplus entries in \texttt{usableAtoms} beyond what the channel will ultimately consume are dropped from the list and added to \texttt{unusableAtoms} for removal.
  \item \textbf{Scatter / Park}: Atoms not needed in the immediate fill are routed to parking sites outside the computation zone for later use; concurrently, \texttt{unusableAtoms} entries are evicted from the working region completely.
  \item \textbf{Fill}: The remaining usable atoms are matched to outstanding target sites and transferred via corridor-routed three- or four-step elbow moves.
\end{itemize}

The Move Encoder shown in the datapath panel of Fig.~\ref{fig:atom-rearrangement-kernel} is unrolled with \texttt{\#pragma HLS UNROLL factor=2}, doubling the effective scatter/fill bandwidth at the cost of a second comparator and writeback port pair. As atoms reach their destinations, the corresponding entries are removed from \texttt{usableAtoms} and \texttt{targetSites} via in-place compaction; the dual-port BRAM allows the read pointer and write pointer to advance in the same cycle. If \texttt{targetSites} is fully drained at the end of S2, the deficiency-resolution stage S3 is skipped entirely, saving one full FSM traversal.

\subsubsection*{Deficiency Resolution (S3)}
This stage is invoked only when some target sites remain unfilled after channel sorting. For each index with unfilled target sites, an index with surplus \texttt{usableAtoms} in parking spots is selected, and the missing atoms are routed using corridor-routed four-step double-elbow moves that traverse the cleared boundary corridor before descending to the destination channel. The control logic for these donor--recipient transfers reuses the same datapath as S2; only the address-generation logic differs, which is why \texttt{FUNCTION\_INSTANTIATE} on the \texttt{vertical} flag is sufficient to specialize the hardware. 

\paragraph{Streaming output to the \ac{AWG}}
Rather than accumulating an off-chip move list, the rearrangement core emits each \emph{ParallelMove} directly onto an AXI-Stream as soon as it is finalized (the ``Moves in Stream'' arrow at the bottom of Fig.~\ref{fig:atom-rearrangement-kernel}, which feeds the AXI-Stream FIFO on the right edge). Each move is packed into a 320-byte packet of up to four steps, where each step carries up to 16 column and 16 row selections as 16-bit signed integers together with their per-axis counts. The number of active steps determines the move type: direct two-step ejection moves for initial corridor clearing, three-step single-elbow moves for in-channel removal moves transfers, and four-step double-elbow moves for cross-channel routing. \texttt{tlast} is asserted on every fifth beat so that one packet exactly delimits one move on the wire.

The AXI-Stream FIFO decouples the producer-side timing of the rearrangement core from the consumer-side \ac{AWG} hardware, absorbing the burstiness of move generation into a smooth output stream. Because moves leave the FPGA in execution order and without an intermediate DDR write-back, the \ac{AWG} can begin acting on the earliest moves while the core is still resolving later ones, overlapping computation and waveform delivery.

%% file: chapters/evaluation.tex
\section{Evaluation}
Having shown the architecture of our system, we can investigate its correctness and performance. We begin by validating the output, followed by an analysis of latency, resource utilization, and scalability.

\subsection{Verification}
In our previous work on atom detection, we have analyzed the algorithm's precision by using a simulated image with known ground-truth occupancy information as the input image \cite{Winklmann:1}. We can now extend this workflow to also include the rearrangement. In summary, we 
\begin{enumerate}
    \item Generate a simulated image based on known occupancy (ground truth).
    \item Conduct atom detection on the simulated image to acquire reconstructed atom positions.
    \item Use reconstructed positions and desired target geometry as input to rearrangement module.
    \item Apply returned moves to initial ground-truth array.
    \item Compare resulting sorted array with target geometry. In practice, one would perhaps acquire another fluorescence image here to detect any errors that occurred due to wrong detection or atoms lost in transit.
\end{enumerate}

\begin{figure}[tb]
  \centering
  \setlength{\fboxsep}{0pt}
  \subfloat[Initial random array]{\fbox{\includegraphics[height=0.33\linewidth]{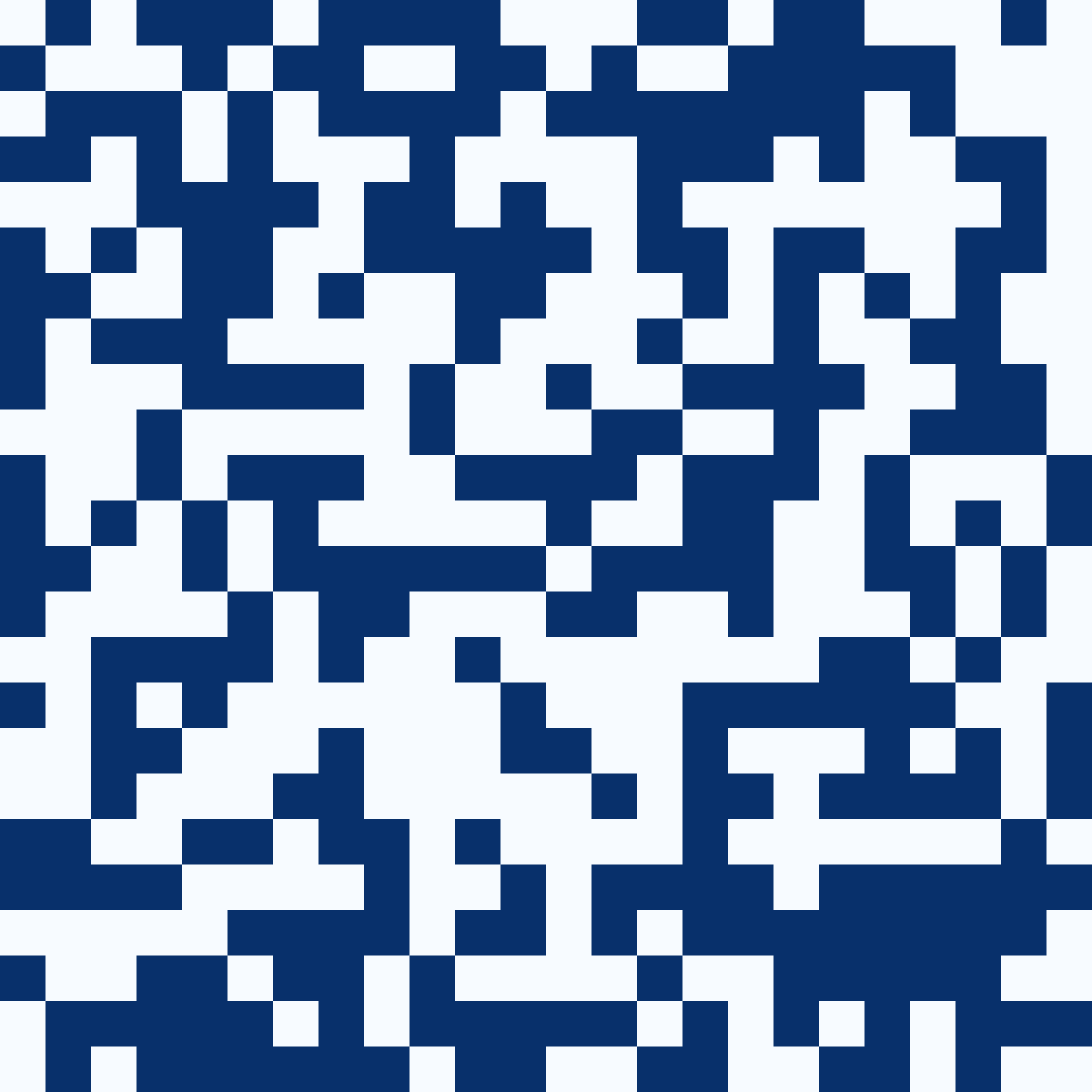}}
  \label{fig:initial_array}}
  \hfill
  \subfloat[Sorted array]{\fbox{\includegraphics[height=0.33\linewidth,]{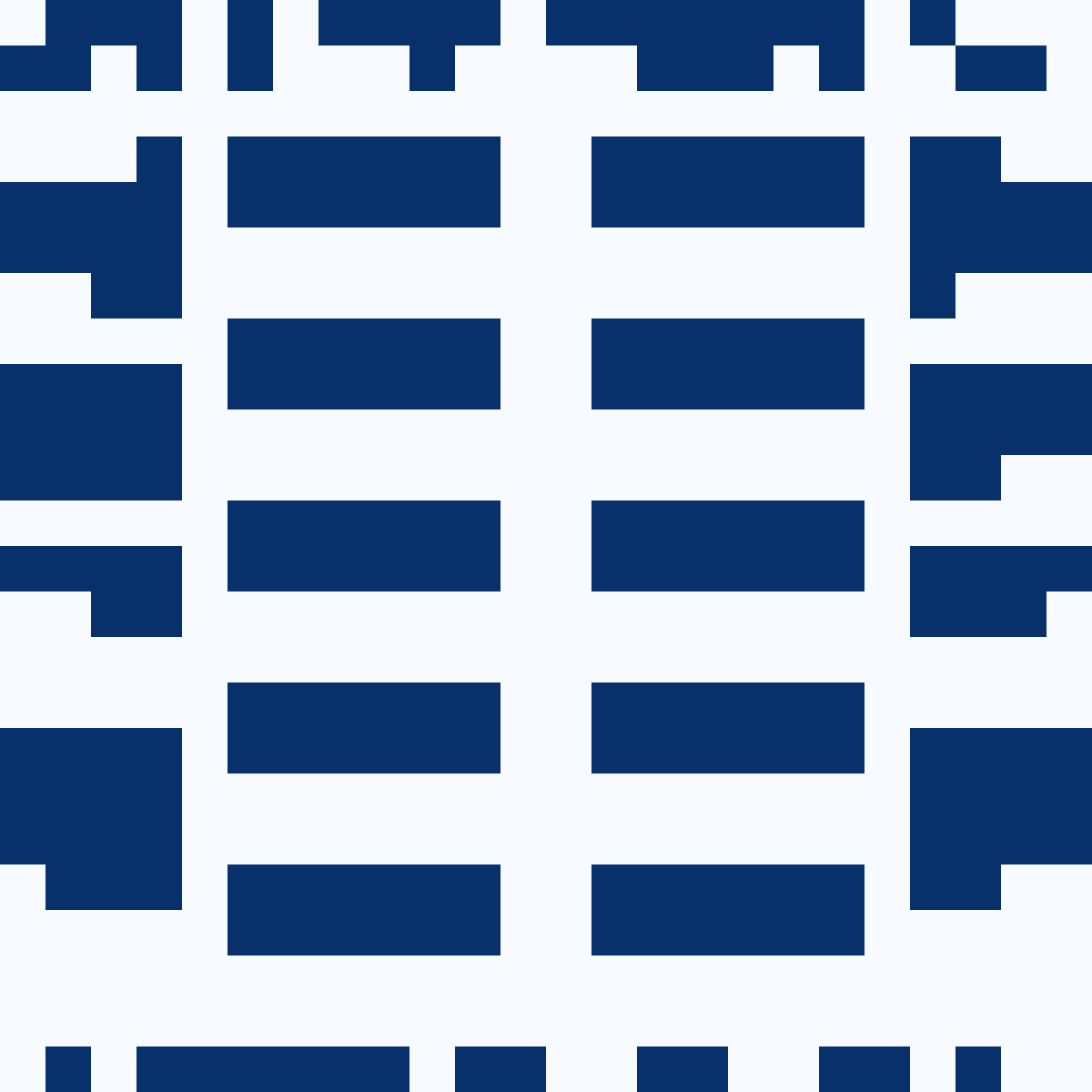}}
  \label{fig:sorted_array}}
  \hfill
  \subfloat[Target geometry]{\fbox{\includegraphics[height=0.33\linewidth,]{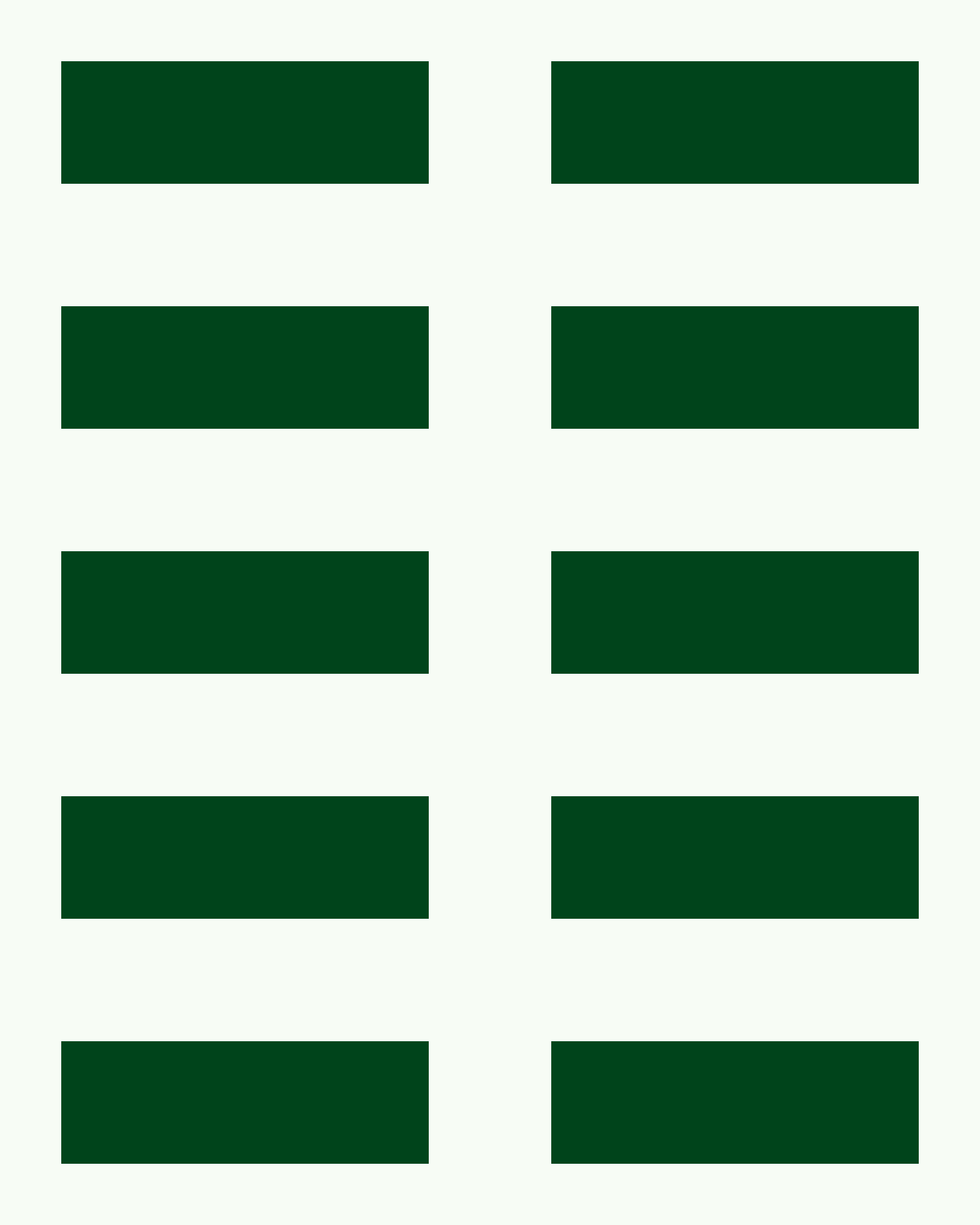}}
  \label{fig:target_geometry}}
  \caption{Visualization of 24$\times$24 rearrangement. (a) Initial random atom array. (b) Final array after rearrangement. (c) Target geometry of size 16$\times$20. It describes for each site whether it will be occupied after sorting or not. The target geometry array is smaller than the total atom array and, in this case, centered within it.}
  \label{fig:sorting_visualization}
\end{figure}

Fig.~\ref{fig:sorting_visualization} shows how applying the returned moves on the initial boolean array yields a central region that matches the provided target geometry. Fig.~\ref{fig:initial_array} shows the ground-truth data, Fig~\ref{fig:sorted_array} shows the array after having applied the moves returned by the rearrangement module, and Fig.~\ref{fig:target_geometry} shows the provided target geometry, matching the center of the actual array after rearrangement. Along the border regions of Fig~\ref{fig:sorted_array}, we can see storage areas where additional atoms are stored. The algorithm may rely on having such parking spots to source atoms from if there is a shortage at the end. Having performed this complete validation workflow, as well as standalone tests on the rearrangement module, we see that no mistakes are introduced through the rearrangement procedure, as long as there is a sufficient number of atoms and, in some cases, parking spots.

\subsection{Latency}
To evaluate the end-to-end performance of the AtomFlow controller, we measured its latency on a Zynq UltraScale+ RFSoC (ZCU216)~\cite{AMDrfsoc} running a representative test case: a $16\times16$ atom array requiring 22 rearrangement moves to reach the target configuration. Each latency breakdown was averaged over 10 independent runs to ensure statistical reliability. The pipeline consists of three main stages: (i) \textit{image analysis}, timed from \texttt{AP\_START} to \texttt{AP\_DONE} in MODE\_QUBIT\_READOUT, which computes per-site emissions from the camera image; (ii) \textit{control interface}, which covers the binary classification of atom occupancy (derived from emissions and an adaptive threshold), FPGA register configuration, and AXI-Lite handshake overhead between the PS and the controller IP; and (iii) \textit{rearrangement}, timed from \texttt{AP\_START} in MODE\_INITIALIZATION until the final move packet is drained from the AXI-Stream FIFO, minus the image analysis and control interface contributions. Each rearrangement move is serialized as a single 320-byte AXI-Stream packet, allowing the PS to decode moves incrementally while sorting is still in progress. The \textit{first-move latency} is defined as the interval between \texttt{AP\_START} and the arrival of the first move packet at the PS, capturing the minimum time before the downstream hardware can begin applying rearrangement pulses.
\begin{table}[htbp]
\centering
\caption{Latency breakdown of the AtomFlow ($16\times16$ atom array, 22 moves)}
\label{tab:latency_breakdown}
\begin{tabular}{lrr}
\toprule
\textbf{Stage} & \textbf{Latency (ms)} & \textbf{\% of Total} \\
\midrule
Image analysis              & 2.4 $\pm$ 0.0   & 9.5\%  \\
Control Interface           & 0.6   & 2.4\%  \\
Rearrangement  & 22.3 $\pm$ 0.2  & 88.1\% \\
\quad \textit{per move}     & \textit{1.01}  & ---    \\
\midrule
\textbf{Total}              & \textbf{25.3 $\pm$ 0.2} & \textbf{100.0\%} \\
\textbf{First-move latency} & \textbf{4}   & ---    \\
\bottomrule
\end{tabular}
\end{table}

Table~\ref{tab:latency_breakdown} summarizes the measured latency breakdown. Image analysis accounts for 2.4~ms (9.5\% of total), and the control interface contributes a small 0.6~ms overhead (2.4\%). Rearrangement is the dominant stage at 22.3~ms (88.1\%), averaging 1.01~ms per move after the first move is generated. The overall end-to-end latency is 25.3~ms, and the first move becomes available to the PS only 4~ms after \texttt{AP\_START}. We utilize the interleaved pipeline mechanism where the host fetches move packets concurrently with FPGA rearrangement, rather than waiting for \texttt{AP\_DONE}. This low first-move latency is critical for hiding rearrangement time behind downstream optical control.

In order to evaluate whether the average time per move is sufficiently fast, we require a model of the time required to execute a given move. With only minor simplification, we describe this time by a constant aspect for transferring between \ac{SLM} and movable \ac{AOD} traps, which we call $t_{transfer}$, and a constant speed $v$ by which the atom with the longest move distance is shifted. For a composite move, as we use them here, we add the longest traversed distances to yield the total move distance $d_{move}$. Therefore, we describe the total execution time as 
\begin{equation}
    t_{exec} = 2\cdot t_{transfer} + \frac{d_{move}}{v}
\end{equation}
Using typical values for $t_{transfer}$ and $v$, as we describe them in Sec.~\ref{chapter:background_rearrangement}, we establish a faster and a slower model. 
\begin{equation}
    t_{exec,fast} = 120\mu s + \frac{d_{move}}{0.55\frac{\mu m}{\mu s}}
\end{equation}
\begin{equation}
    t_{exec,slow} = 800\mu s + \frac{d_{move}}{0.054\frac{\mu m}{\mu s}}
\end{equation}

\begin{figure}[tb]
   \centering
\includegraphics[width=1\linewidth]{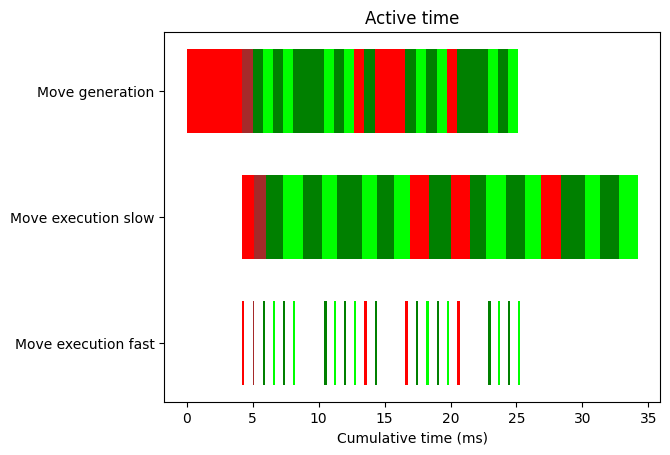}
 \caption{Timeline of active and idle time for move-generation and move-execution modules (based on one fast and one slow model). Move execution can only begin once the next move is generated. Red colors denote moves that remove unusable atoms, green colors are used for all other moves. Color shades alternate to allow for differentiation of moves.}
 \label{fig:arrivalAndExecTimes}
\end{figure}

Fig.~\ref{fig:arrivalAndExecTimes} shows, for a single exemplary rearrangement run on 16$\times$16 atom sites with a separation of 4$\mu$m per site, the time stamps when each move was generated and the time stamp of when physical execution starts and ends using our two execution-time models. Whenever there is a gap in the move-execution timeline, the next move is not yet available, and the waveform generator has to wait. This is not ideal as, in this case, the time taken to generate the next move is not covered by the time it took to execute the previous one. For the slow model, we can see that our solution is fast enough to always create moves in time for them to be executed. In fact, we can see that there are usually excess moves in the output queue. The only latency of our rearrangement module that is relevant to the overall sorting time is therefore the latency of generating the first move. On the other hand, it is clear that our solution needs further optimizations in order to fully hide its run time behind the physical rearrangement time if the fast model is considered.
\subsection{Resources}

Since we want to place several control modules on the same device, resource utilization is relevant. Table~\ref{tab:resource_breakdown} reports the post-implementation FPGA resource utilization for the $16\times16$ atom array test case, extracted from a hierarchical utilization report after place-and-route. The full pipeline fits comfortably within the device, consuming 29.6\% of LUTs, 21.1\% of FFs, 20.8\% of BRAMs, and 11.1\% of DSPs. Image analysis dominates DSP usage (447 of 475) due to the intensive multiplication in the \ac{PSF}-based reconstruction, and also accounts for the largest LUT/FF consumption (87.6k LUTs, 124.2k FFs). Atom rearrangement, despite its algorithmic complexity, uses far fewer LUTs and DSPs than image analysis after implementation optimization, but dominates BRAM usage (202 tiles) for the state arrays and parallel-move buffers. The control interface and AXI infrastructure together take fewer than 20k LUTs. The remaining resources, particularly in DSPs and BRAMs, indicate that the design can be scaled up to larger atom arrays without exceeding device capacity.
\begin{table}[t]
\centering
\caption{FPGA resource utilization for $16\times16$ atom array (post-implementation, extracted from hierarchical report)}
\label{tab:resource_breakdown}
\resizebox{\columnwidth}{!}{%
\begin{tabular}{lrrrr}
\toprule
\textbf{Module} & \textbf{LUT} & \textbf{FF} & \textbf{BRAM(36K)}& \textbf{DSP} \\
\midrule
Image analysis          &  87{,}599 & 124{,}194 &  23   & 447 \\
Control interface       &  10{,}032 &  14{,}532 &   0   &  17 \\
Atom rearrangement      &  18{,}567 &  23{,}398 & 22   &  11 \\
AXI \& I/O buffers      &   9{,}838 &  17{,}149 &   1   &   0 \\
\midrule
\textbf{Total}          & \textbf{126{,}036} & \textbf{179{,}273} & \textbf{46} & \textbf{475} \\
\textbf{Utilization}    & 29.6\%    & 21.1\%    & 4.3\% & 11.1\% \\
\bottomrule
\end{tabular}%
}
\end{table}

\subsection{Scalability}
\begin{figure}[tb]
   \centering
\includegraphics[page=1,width=\linewidth]{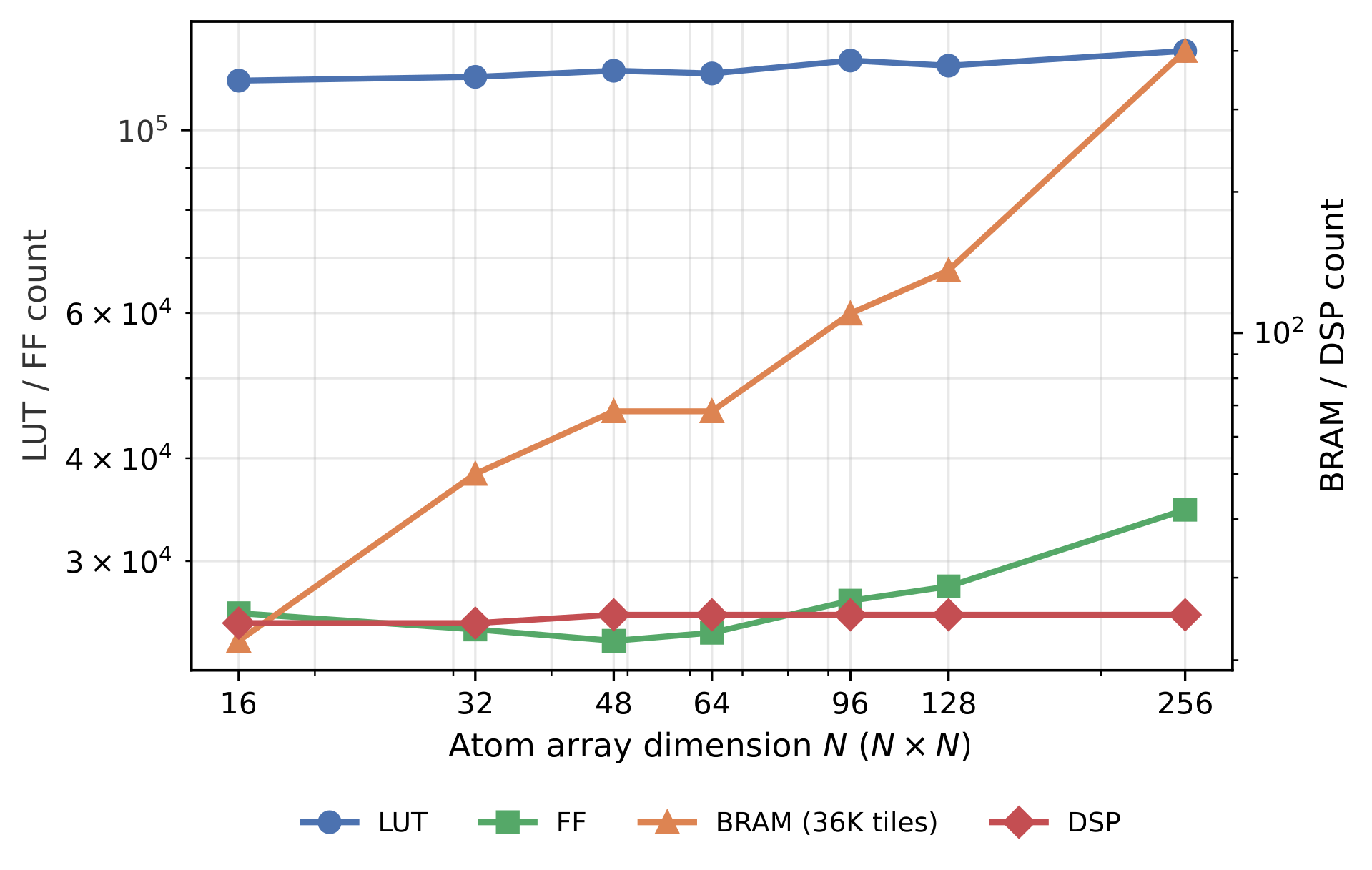}
 \caption{FPGA resource scaling of the atom rearrangement engine (data from HLS synthesis) for atom array dimensions $N = 16$ to $256$. Note that these values are reported directly from \ac{HLS} synthesis estimates; the actual post-implementation utilization in Vivado is typically substantially lower due to backend optimizations (e.g., LUT combining, constant propagation), so the figures shown here represent a conservative upper bound.}
 \label{fig:scaling}
\end{figure}

Our previous work~\cite{jonas_dac} has already conducted a scalability analysis of the \textit{image analysis module}. Moreover, the experimental setup required for \textit{image analysis} is significantly smaller than that of the \textit{rearrangement engine}, which prevents a unified scalability study covering both stages. We therefore focus here exclusively on the scalability of atom rearrangement. Fig.~\ref{fig:scaling} illustrates the change of FPGA resources when we vary the atom array dimension from 16 to 256. For each data point for the array dimension $N$, we proportionally scale the initial atom array as well as the target geometry (grid state buffers), while keeping the system-wise parameters (parallel moves) unchanged, since the latter one depends on the physical \ac{AOD} limits instead of a scalable factor.

We observe from the figure that only BRAM usage grows approximately linearly with $N$, due to the $N\times N$ configuration of the internal state buffers. Notably, because BRAM blocks have the quantization behavior, especially at small $N$, it flattens the otherwise $N^2$ scaling. On the contrary, LUT, FF, and DSP remain nearly stable, since the combinational logic of the rearrangement pipeline is dominated by fixed-width arithmetic over the bounded AOD channel array, independent of grid size. It should be noted that the resource counts reported in Fig.~\ref{fig:scaling} are pre-optimization \ac{HLS} estimates and thus represent only an upper bound. The deployed utilization after Vivado place-and-route is commonly substantially lower. Accordingly, the focus of this analysis is on the \textit{scaling trend} across atom-array dimensions rather than on absolute resource counts. Together with Table~\ref{tab:resource_breakdown}, these results demonstrate that the \textit{rearrangement engine} scales well within current device capacity. Even if BRAM demand exceeds the on-chip budget at larger array sizes, the caching strategy adopted in our \textit{image-analysis module} can be applied to substantially reduce BRAM consumption.

%% file: chapters/conclusion.tex
\section{Conclusion}
In this project, we aimed at developing a fully integrated atom-detection and rearrangement pipeline called \textit{AtomFlow} with a focus on the interfaces and communication in order to minimize the overall latency of the initialization procedure of \acp{NAQC}. To do so, we first had to adapt a suitable rearrangement algorithm for use on \acp{FPGA}, which we did using \ac{HLS}. The resulting module supports up to 256$\times$256 atom sites, can deal with various constraints to allow for usage in different hardware setups, and has a sufficiently low resource utilization to be combined with our other solutions. We integrated this module with our previous work on \ac{FPGA}-based image analysis for atom detection, which allows us to fully close the loop by connecting directly to the camera and waveform generation through low-level interfaces.

For future work, there remain some shortcomings that we need to remedy. In terms of scaling to higher qubit numbers, the image analysis currently represents the bottleneck, although we do not expect major issues in scaling beyond its current scalability limits. Regarding our solution's latency, there is room for improvement on the rearrangement module. Even though nearly the whole run time can be hidden behind the physical move-execution times for conservative models, further maturation of the technology is to be expected, carrying with it improved transfer times between \ac{AOD} and \ac{SLM} traps, as well as optimized acceleration curves, which could hugely benefit from faster move generation.